\documentstyle[aps,preprint,epsf]{revtex}

\newcommand{\be}{\begin{eqnarray}}
\newcommand{\ee}{\end{eqnarray}}
\begin{document}

\title{A linked cluster expansion for the calculation of the semi-inclusive  
$A(e,e'p)X$ processes using correlated Glauber wave functions}
\draft

\author{Claudio\,Ciofi degli Atti}

\address{ Department of Physics, University of Perugia, \\and \\Istituto Nazionale 
di Fisica Nucleare, Sezione di Perugia, \\Via A. Pascoli, I--06100 Perugia, Italy}
\author{Daniele\, Treleani}

\address{ Department of Theoretical Physics, University of Trieste, \\and\\ Istituto Nazionale di Fisica Nucleare, Sezione di Trieste,
and ICTP, \\Strada
Costiera 11, I--34014, Trieste,Italy}
\date{\today}

\maketitle

\begin{abstract}
 The distorted one-body mixed density matrix, which is the basic nuclear quantity appearing in the definition of the cross section for the semi-inclusive $A(e,e'p)X$ processes, is calculated within a linked-cluster expansion based upon  correlated wave functions and the Glauber multiple scattering theory to take into account the final state interaction of the ejected nucleon. The nuclear transparency for $^{16}O$ and $^{40}Ca$ is calculated using realistic 
central and non-central correlations and the important role played by the latter is illustrated.
\end{abstract}


\section{Introduction}

The accurate calculation of the final state interactions (FSI) of the ejected nucleons in exclusive and semi-inclusive processes of the type $A(e,e'N)(A-1)$, 
$A(e,e'N)X$, $A(e,e'NN)X$ etc. induced by medium- and high-energy electrons, 
is one of the most urgent and important theoretical challanges in the investigation of the properties of hadronic matter. As a matter of fact, the possibilities to get information on basic properties
of bound hadrons, such as, for example  their  momentum and energy distributions, crucially depend upon the ability to estimate to which extent
FSI effects  destroy the direct link between the measured  cross section and the hadronic properties before interaction with the probe, which is generally provided by approximations, e.g. the impulse approximation (IA), which disregard FSI  (see e.g. \cite{boffi}). Another convincing motivation for
an accurate treatment  of FSI, stems from the expectation that at large $Q^2$ they should vanish because of  Color Transparency (CT), an effect  originally predicted  by
Brodsky \cite{brodsky} and Mueller \cite{mueller},  and extensively investigated by various authors (for  recent reviews on the subject, see e.g. \cite{color}), according to which the ejectile rescattering amplitudes with elastic and
inelastic intermediate states interfere destructively. Since the onset of the phenomenon is expected to show up at large  values of $Q^2$, when  
FSI effects could be evaluated within the standard Glauber theory, 
the experimental investigation of CT relies on the detection of possible differences between  experimental data and predictions of standard Glauber multiple scattering calculations of FSI. However, due to the expected small difference, an accurate treatment of nuclear structure effects is a prerequisite in order to get reliable informastion on CT effects. Among the  large variety of nuclear  effects,  those produced  by nucleon-nucleon (NN) correlations, which will be called from now on initial state correlations (ISC), play a  dominant  role, for many-body calculations based upon realistic NN interaction models predict a rich correlation structure of the nuclear wave function (see e.g. \cite{manybody}. The effect of NN correlations in the calculation of   FSI
within the Glauber approach, have been considered in various papers \cite{seki} - \cite{benhar}, where, due to the difficulty of the problem, various approximations have been introduced either by  truncating  the Glauber multiple scattering series, or by considering oversimplified models of correlations, e.g. by adopting simple phenomenological Jastrow-type wave functions embodying only central correlations.
 
In  this paper  a novel approach to the problem is presented, 
based upon a 
linked-cluster expansion series of the distorted one-body mixed density matrix starting 
from realistic correlated wave functions and Glauber multiple scattering 
operators. The expansion is such that, at each order in the correlations, Glauber multiple scattering is included at all order. The expansion  is based upon the number conserving approach of (\cite{ripka}), properly generalised to take into account Glauber FSI. 
  
Our paper is organised as follows: the basic elements of 
the theory i.e. the concepts of semi-inclusive processes $A(e,e'N)X$, nuclear transparency and distorted momentum distributions  are reviewed in Section II; the formal developments of the  linked-cluster  
expansion are illustrated in Section III; the basic elements underlying the 
calculations of the nuclear transparency, i.e. the correlated nuclear wave 
function
and the Glauber multiple scattering operators are discussed in section IV, where 
the results of the calculations of the nuclear transparency in the processes
$^{16}O(e,e'p)X$ and $^{40}Ca(e,e'p)X$ are also presented; finally, the Summary and Conclusions are 
given in Section V.

\section{The  semi-inclusive process {\bf\it A(\lowercase{e,e'p})X}, the nuclear transparency and the distorted momentum distributions }

We will consider the process $A(e,e'p)X$ in which an electron with 4-momentum
$k_1\equiv\{{\bf k}_1,i\epsilon_1\}$, is scattered off a nucleus with 4-momentum
$P_A\equiv\{{\bf 0},iM_A\}$ to a state $k_2\equiv\{{\bf k}_2,i\epsilon_2\}$ 
and is detected in coincidence with a proton $p$ with 4-momentum 
$k_p\equiv\{{\bf k}_p,iE_p\}$; the final $(A-1)$ nuclear system with 
4-momentum $P_X\equiv\{{\bf P}_X,iE_X\}$ is undetected. The cross section 
describing the process can be written as follows

\begin{equation}
\frac{d\sigma}{dQ^2d\nu d{\bf k}_p}=K\sigma_{ep}P_D(E_m,{\bf k}_m)
\label{sezione}
\end{equation}
where $K$ is a kinematical factor, $\sigma_{ep}$ the off-shell electron-nucleon 
cross section,  and $Q^2=|{\bf q}|^2-\nu^2$ the four momentum transfer. The 
quantity $P_D(E_m,{\bf k}_m)$ is the distorted nucleon spectral function which 
depends upon the observable {\it missing momentum} 

\begin{equation}
{\bf k}_m={\bf q}-{\bf k}_p
\label{missingmom}
\end{equation}
and {\it missing energy}

\begin{equation}
E_m=\nu+M-E_p
\label{missingen}
\end{equation}
The latter equation results from energy conservation

\begin{equation}
\nu+M_A=E_p+\sqrt{M_X^2+{\bf p}_X^2}
\label{energycons}
\end{equation}
if the total energy of the system $X$ is
approximated by its non-relativistic
expression and the recoil energy is disregarded. The distorted spectral function 
can be written in the following short-hand form \cite {nikolayev}

\begin{equation}
P_D(E_m,{\bf k}_m)=\sum_{f_X}|\langle{\bf 
k}_m,\Psi_{f_X}|\Psi_A\rangle|^2\delta(E_m-(E_{min}+E_{f_X}))
\label{Pempm}
\end{equation}

where $E_{min} = M+M_{A-1}-M_A$, and 

\begin{equation}
\langle{\bf k}_m,\Psi_{f_X}|\Psi_A\rangle=\int e^{i{\bf k}_m{\bf r}_1}S_G^{\dagger}({\bf 
r}_1\dots{\bf r}_A)\Psi_{f_X}^*({\bf r}_2\dots{\bf r}_A)\Psi_A({\bf 
r}_1\dots{\bf r}_A)\delta ({\sum_{j=1}^A {\bf r}_j})\prod_{i=1}^Ad{\bf r}_i,
\label{overlap}
\end{equation}
with $\Psi_A$ and $\Psi_{f_X}$ being the ground state wave 
function of the target nucleus and the wave function of the system $X$ in the 
state $f_X$, respectively; the quantity $S_G$ is the Glauber operator, which describes the  FSI of the struck proton with the $(A-1)$ system, i. e. 

\begin{equation}
S_G({\bf r}_1\dots{\bf r}_A)=\prod_{j=2}^AG({\bf r}_1,{\bf r}_j)\equiv
\prod_{j=2}^A\bigl[1-\theta(z_j-z_1)\Gamma({\bf b}_1-{\bf b}_j)\bigr]
\label{SG}
\end{equation}
where ${\bf b}_j$ and $z_j$ are the transverse and the longitudinal components 
of the nucleon coordinate ${\bf r}_j\equiv({\bf  b}_j,z_j)$, ${\mit\Gamma}({\bf 
b})$ is
the Glauber profile function for elastic proton nucleon scattering, and the 
function $\theta(z_j-z_1)$ takes care of the fact that the struck proton ``1'' 
propagates along a straight-path trajectory so that it interacts with nucleon 
``$j$'' only if $z_j>z_1$. The integral over the missing energy of the distorted 
spectral function defines the distorted momentum distribution as

\begin{equation}
n_D({\bf k}_m)=\int dE_m P_D(E_m,{\bf k}_m)
\label{nd}
\end{equation}

In impulse approximation (IA) (i.e. when the final state interaction is 
disregarded ($S_G=1$)), if the system $X$ is assumed to be a $(A-1)$ nucleus in 
the discrete or continuum states $f_X\equiv f_{A-1}$, the distorted spectral function 
$P_D$ reduces to the usual spectral function, i.e.

\begin{equation} 
P_D\to P(k,E)=\sum_{f_{A-1}}|\langle{\bf 
k},\Psi_{f_{A-1}}|\Psi_A\rangle|^2\delta\bigl(E-(E_{min}+E_{f_{A-1}})\bigr
)
\label{Pke}
\end{equation}
where $E$ is the nucleon removal energy i.e. the energy required to remove a 
nucleon from the target, leaving the $A-1$ nucleus with excitation energy 
$E_{f_{A-1}}$ and ${\bf k}=-{\bf k}_m={\bf q}-{\bf k}_p$ is the nucleon momentum 
before interaction. The integral of the spectral function over the $E$ defines the (undistorted)
momentum distributions 
\begin{equation}
n({\bf k})=\int dE P(E,{\bf k})
\label{nk}
\end{equation}

In this paper we will consider the effect of the FSI ($S_G\not=1$) on the 
semi-inclusive $A(e,e'p)X$ process, i.e.  the cross section (\ref{sezione})
integrated over the missing energy $E_m$, at fixed value of $\bf p_m$. Owing to 

\begin{equation}
\sum_{f_X}\Psi_{f_X}^*({\bf r}_2'\dots{\bf r}_A')\Psi_{f_X}({\bf r}_2\dots{\bf 
r}_A)=\prod_{j=2}^A\delta({\bf r}_j-{\bf r}_j')
\label{closure}
\end{equation}
the cross section (\ref{sezione}) becomes directly proportional to the distorted 
momentum distributions (\ref{nd}), i.e. 

\begin{equation}
n_D({\bf k}_m)={(2 \pi)^{-3}} \int e^{i {\bf k}_m({\bf r} -{\bf r}')}\rho_D 
({\bf r},{\bf r}') d{\bf r} d{\bf r}'
  \label{nd1}
   \end{equation}
where 
\begin{eqnarray}
\rho_D ({\bf r},{\bf r}')= \frac {\langle\Psi_A S_G^{\dagger} \hat{O}({\bf 
r},{\bf r}')  S_G'{\Psi_A}'\rangle}{\langle\Psi_A\Psi_A\rangle}
  \label{rodi}
   \end{eqnarray}
is the  one-body mixed density matrix, and
\begin{eqnarray}
\hat{O}({\bf r},{\bf r}')= \sum_i\delta({\bf r_i} - {\bf r}) \delta(\bf r_i^{'} - \bf r^{'})\prod_{j\not= i}\delta({\bf r_j}-{\bf r_j}')
\label{rodiop}
   \end{eqnarray}
the one-body density operator.
In Eq. (\ref{rodi}) and in the rest of the paper, the primed quantities have to 
be evaluated at ${\bf r}'$ with $i=1, ...,A$. By integrating $n_D({\bf k}_m)$
  the nuclear transparency $T$ is obtained, which is defined as follows

\begin{eqnarray}
T=\int n_D({\bf k}_m) d{\bf k}_m = (2 \pi)^{-3}\int \rho_D ({\bf r},{\bf r}') d{\bf r} d{\bf 
r}'
\int e^{i {\bf k}_m({\bf r} -{\bf r}')}d{\bf k}_m = \int \rho_D ({\bf r})d {\bf 
r}
  \label{intnd}
   \end{eqnarray}
i.e.

\begin{equation}
T = \int\rho_D ({\bf r})d {\bf r} = 1+ \Delta T
  \label{ti}
   \end{equation}
where $\Delta T$ originates from FSI.
The nuclear momentum distributions and the one-body density are normalised as follows

\begin{eqnarray}
\int n({\bf p}) d{\bf p} = \int \rho ({\bf r})d {\bf r} = 1
  \label{norma}
   \end{eqnarray}
   
\section{The one-body mixed density matrix and nuclear transparency with a 
linked cluster expansion for Glauber correlated wave functions}

We  have evaluated the one-body density matrix (\ref{rodi}) using for $S_G$ the 
form ( \ref{SG})
and for the nuclear wave function $\Psi_{A}$ the following form

\begin{equation}
\Psi_A = {\rm {\hat{S}}} \Big [ \prod_{i<j}\hat {f}(ij)\Big] 
 \Psi_0
  \label{psi}
   \end{equation}
where
\begin{equation}
\hat{f}(ij)=\sum_n {f_n (r_{ij})\hat {O}_n (ij)}
\label{effe}
\end{equation}
$\rm {\hat{S}}$ is the symmetrisation operator, $\Psi_0$   the Slater 
determinant describing the nucleon independent particle motion, and $f_n 
(r_{ij})$ the correlation function associated to the operator $\hat{O}_n (ij)$ 
(if $\hat{O}_n (ij) = 0$ for $n>1$, the usual Jastrow wave function is recovered).
If Glauber FSI and nucleon-nucleon correlations are both absent ($S_G=1$, $f_1=1$, $f_n=0$, for 
$n>1$) the standard  results for the shell-model one-body mixed density matrices: 

\begin{equation}
\rho_{SM}({\bf r}, {\bf r}')=\sum_{\alpha}\phi_{\alpha}^*({\bf 
x})\phi_{\alpha}({\bf x}')=4\rho_0({\bf r}, {\bf r}')
\label{rorr'}
\end{equation}

\begin{equation}
\rho_{SM}({\bf r}_i, {\bf r}_j)=\sum_{\alpha}\phi_{\alpha}^*({\bf 
x}_i)\phi_{\alpha}({\bf x}_j)=4\rho_0({\bf r}_i, {\bf r}_j)
\label{roij}
\end{equation}
and the one-body diagonal matrix

\begin{equation}
\rho_{SM}({\bf r}_i)\equiv\rho_{SM}({\bf r}_i, {\bf 
r}_i)=\sum_{\alpha}|\phi_{\alpha}({\bf x}_i)|^2=4\rho_0({\bf r}_i)
\label{roi}
\end{equation}
are obtained, where 

\begin{equation}
\rho_0({\bf r}_i, {\bf r}_j)=\sum_a\varphi^*_a({\bf r}_i)\varphi_a({\bf r}_j)
\label{rorirj}
\end{equation}
and
\begin{equation}
\rho_0({\bf r}_i)=\sum_a|\varphi_a({\bf r}_i)|^2
\label{rori}
\end{equation}
are the spin- and isospin-independent matrices. In the above equations,  
 the notations
$\alpha\equiv\{a,\sigma,\tau\}$, $a\equiv\{n,l,m\}$, and ${\bf
x}\equiv\{{\bf r}, {\bf s}, {\bf t}\}$, have been used, which means that the single particle orbitals have the following representation $\phi_{\alpha}({\bf
x})=\varphi_a({\bf r}){\bf \chi}_{\sigma}^{1/2}\xi_{\tau}^{1/2}$.

We have developed a linked cluster expansion in the  quantity
$\eta ({\bf r}_i,{\bf r}_j,{\bf r}_i',{\bf r}_j')= 1 + f^*({\bf r}_i,{\bf 
r}_j)f({\bf r}_i',{\bf r}_j')$ which includes, at each order in
$\eta({\bf r}_i,{\bf r}_j,{\bf r}_i',{\bf r}_j')$, the Glauber operator to all orders, and the  result at first order reads as follows

\begin{eqnarray}
\rho_D ({\bf r}_1,{\bf r}_1') &\simeq &<\Psi_o \mid \prod G^{\dagger}({\bf 
r}_1,{\bf r}_i) { \hat O }({\bf r}_1,{\bf r}_1')\prod G({\bf r}_1',{\bf 
r}_j')\mid \Psi_o'>\nonumber\\ &+&
 <\Psi_o \mid \prod G^{\dagger}({\bf r}_1,{\bf r}_i) \sum \eta({\bf r}_i,{\bf 
r}_j,{\bf r}_i',{\bf r}_j'){ \hat O }({\bf r}_1,{\bf r}_1')\prod G({\bf 
r}_1',{\bf r}_j')\mid \Psi_o'>\nonumber\\ &-&
<\Psi_o \mid \sum\eta({\bf r}_i,{\bf r}_j,{\bf r}_i,{\bf 
r}_j)\mid\Psi_o><\Psi_o\mid\prod G^{\dagger}({\bf r}_1,{\bf r}_i){ \hat O }({\bf r}_1,{\bf r'}_1)G({\bf 
r'}_1,{\bf r'}_i) \mid \Psi_o'>\
  \label{rodLOC}
   \end{eqnarray}
Placing Eq. (\ref{rodiop}) in the above equation, one obtains

\begin{equation}
\rho_D ({\bf r}_1,{\bf r}_1')= 
\tilde{A}+\tilde{B}_1+\tilde{B}_2^L+\tilde{B}_2^U-\tilde{C}^U-\tilde{C}^L
 \label{rodLOC1}
   \end{equation}
where
\begin{equation}
\tilde{A}=\rho_{SM}({\bf r}_1,{\bf r}_1')\times\Phi({\bf r}_1,{\bf r}_1')^{(A-1)}
\label{atilde}
\end{equation}

\begin{eqnarray}
\tilde{B}_1=4\Phi({\bf r}_1,{\bf r}_1')^{(A-2)}\int d{\bf r}_2\Bigl\{ 
\bigl[4&H&^{dir}(r_{12},r_{1'2})
\rho_0({\bf r}_1,{\bf r}_1')\rho_0({\bf r}_2)\nonumber\\
-&H&^{ex}(r_{12},r_{1'2})\rho_0({\bf r}_1,{\bf r}_2)\rho_0({\bf r}_2,{\bf 
r}_1')\bigr]
G^{\dagger}({\bf r}_1,{\bf r}_2)G({\bf r}_1',{\bf r}_2)\Bigr\}
\label{b1tilde}
\end{eqnarray}

\begin{eqnarray}
\tilde{B}_2^L=&-&4\Phi({\bf r}_1,{\bf r}_1')^{(A-3)}\sum_{a\not= 
b}\varphi_a^*({\bf r}_1)\varphi_b({\bf r}_1')\int d{\bf r}_2d{\bf r}_3\Bigl\{ 
\bigl[4H^{dir}(r_{23})
\varphi_b^*({\bf r}_2)\varphi_a({\bf r_2})\rho({\bf r}_3)\nonumber\\
&-&H^{ex}(r_{23})
\varphi_b^*({\bf r}_2)\varphi_a({\bf r}_3)\rho({\bf r}_3,{\bf r}_2)\bigr]
G^{\dagger}({\bf r}_1,{\bf r}_2)G^{\dagger}({\bf r}_1,{\bf r}_3)
G({\bf r}_1',{\bf r}_2)G({\bf r}_1',{\bf r}_3)\Bigr\}
\label{b2tildeL}
\end{eqnarray}

\begin{eqnarray}
\tilde{B}_2^U=&4&\Phi({\bf r}_1,{\bf r}_1')^{(A-3)}\sum_{a\not= 
b}\varphi_a^*({\bf r}_1)\varphi_a({\bf r}_1')\int d{\bf r}_2d{\bf r}_3\Bigl\{ 
\bigl[4H^{dir}(r_{23})
|\varphi_b({\bf r}_2)|^2\rho({\bf r}_3)\nonumber\\
&-&H^{ex}(r_{23})
\varphi_b^*({\bf r}_2)\varphi_b({\bf r}_3)\rho_0({\bf r}_3,{\bf r}_2)\bigr]
G^{\dagger}({\bf r}_1,{\bf r}_2)G^{\dagger}({\bf r}_1,{\bf r}_3)
G({\bf r}_1',{\bf r}_2)G({\bf r}_1',{\bf r}_3)\Bigr\}
\label{b2tildeU}
\end{eqnarray}

\begin{eqnarray}
\tilde{C}^L=&4&\Phi({\bf r}_1,{\bf r}_1')^{(A-1)}\sum_{a}\varphi_a^*({\bf 
r}_1)\varphi_a({\bf r}_1')\int d{\bf r}_2d{\bf r}_3\Bigl\{ 
\bigl[4H^{dir}(r_{23})
|\varphi_a({\bf r}_2)|^2\rho_0({\bf r}_3)\nonumber\\
&-&H^{ex}(r_{23})
\varphi_a^*({\bf r}_2)\varphi_a({\bf r}_3)
\rho_0({\bf r}_3,{\bf r}_2)\bigr]\Bigr\}
\label{ctildeL}
\end{eqnarray}

\begin{eqnarray}
\tilde{C}^U=&4&\Phi({\bf r}_1,{\bf r}_1')^{(A-1)}\sum_{a\not=b}\varphi_a^*({\bf 
r}_1)\varphi_a({\bf r}_1')\int d{\bf r}_2d{\bf r}_3\Bigl\{ 
\bigl[4H^{dir}(r_{23})
|\varphi_b({\bf r}_2)|^2\rho_0({\bf r}_3)\nonumber\\
&-&H^{ex}(r_{23})
\varphi_b^*({\bf r}_2)\varphi_b({\bf r}_3)
\rho_0({\bf r}_3,{\bf r}_2)\bigr]\Bigr\}
\label{ctildeU}
\end{eqnarray}
where  $\rho_0({\bf r}_i,{\bf r}_j)$ and $\rho_0({\bf 
r}_i)$  are defined by Eq.'s
(\ref{rorirj}) and (\ref{rori}),respectively,  $H^{dir(ex)}(r_{12},r_{1'2})$ and 
$H^{dir(ex)}(r_{23})$, where  {\it dir(ex)} stand for {\it direct(exchange)}, respectively, depend upon the form of the correlation operator in 
(\ref{psi}) and will be defined in section 4, and

\begin{equation}
\qquad\big [\Phi ({\bf r}_1,{\bf r}_1')\big]^n\equiv\Big [\int \rho_o({\bf r}_j) 
G^{\dagger}({\bf r}_1,{\bf r}_j)G({\bf r}_1',{\bf r}_j)d{\bf r}_j\Big]^n
\label{figlauber}
\end{equation}
with $n =(A-3), (A-2), (A-1), A$. In the above equations the sum over $a$ and $b$ runs over shell model occupied 
states below the Fermi sea.

Equation (\ref{rodLOC}) holds for any value of $A$. We will now consider the 
usual Glauber condition
 of large $A$, i.e. we consider $n=(A-3)\simeq(A-2)\simeq(A-1)\simeq A$; in such a 
case the various terms of Eq. (\ref{rodLOC}) can be properly rearranged to finally 
obtain the following compact result
 
\begin{eqnarray}
\rho_D ({\bf r}_1,{\bf r}_1') \simeq \rho_{SM}({\bf r}_1,{\bf r}_1') 
&+&\rho_{ISC}^H({\bf r}_1,{\bf r}_1') +\rho_{ISC}^S({\bf r}_1,{\bf r}_1') 
+\rho_{FSI}^{SM}({\bf r}_1,{\bf r}_1') \nonumber\\
&+&\rho_{FSI}^H({\bf r}_1,{\bf r}_1')  +\rho_{FSI}^S({\bf r}_1,{\bf r}_1') 
\label{rodifinale}
   \end{eqnarray}

The physical meaning of the various terms in Eq. (\ref{rodifinale}) will be discussed later on; their explicit form is as follows 

\begin{eqnarray}
\rho_{ISC}^H({\bf r}_1,{\bf r}_1')=4\int d{\bf r}_2\Bigl\{ 
\bigl[4&H&^{dir}(r_{12},r_{1'2})
\rho_0({\bf r}_1,{\bf r}_1')\rho_0({\bf r}_2)\nonumber\\
-&H&^{ex}(r_{12},r_{1'2})\rho_0({\bf r}_1,{\bf r}_2)\rho_0({\bf r}_2,{\bf 
r}_1')\bigr]\Bigr\}
\label{roISCH}
\end{eqnarray}

\begin{eqnarray}
\rho_{ISC}^S({\bf r}_1,{\bf r}_1')=-4\int d{\bf r}_2d{\bf r}_3\Bigl\{ 
\bigl[4&H&^{dir}(r_{23})
\rho_0({\bf r}_2,{\bf r}_1')\rho_0({\bf r}_3)\nonumber\\
-&H&^{ex}(r_{23})\rho_0({\bf r}_2,{\bf r}_3)\rho_0({\bf r}_3,{\bf 
r}_1')\bigr]\rho_0({\bf r}_1,{\bf r}_2)\Bigr\}
\label{roISCS}
\end{eqnarray}

\begin{equation}
\rho_{FSI}^{SM}({\bf r}_1,{\bf r}_1')=\Bigl\{\rho_{SM}({\bf r}_1,{\bf r}_1') 
+\rho_{ISC}^H({\bf r}_1,{\bf r}_1') +\rho_{ISC}^S({\bf r}_1,{\bf r}_1')\Bigr\}
\Bigl\{\Phi ({\bf r}_1,{\bf r}_1')^A-1\Bigr\}
\label{roFSISM}
\end{equation}

\begin{eqnarray}
\rho_{FSI}^H({\bf r}_1,{\bf r}_1')=\Phi ({\bf r}_1,{\bf r}_1')^A\times
4\int d{\bf r}_2\Bigl\{ \bigl[4&H&^{dir}(r_{12},r_{1'2})
\rho_0({\bf r}_1,{\bf r}_1')\rho_0({\bf r}_2)\nonumber\\
-&H&^{ex}(r_{12},r_{1'2})\rho_0({\bf r}_1,{\bf r}_2)\rho_0({\bf r}_2,{\bf 
r}_1')\bigr]\Gamma({\bf r}_1,{\bf r}_1',{\bf r}_2)\Bigr\}
\label{roFSIH}
\end{eqnarray}

\begin{equation}
\rho_{FSI}^{S}({\bf r}_1,{\bf r}_1')=\rho_{FSI}^{S,L}({\bf r}_1,{\bf r}_1')+
\rho_{FSI}^{S,U}({\bf r}_1,{\bf r}_1')
\label{roFSIS}
\end{equation}
with

\begin{eqnarray}
\rho_{FSI}^{S,L}({\bf r}_1,{\bf r}_1')=-\Phi ({\bf r}_1,{\bf r}_1')^A&\times&
4\int d{\bf r}_2d{\bf r}_3\Bigl\{ \bigl[4H^{dir}(r_{23})
\rho_0({\bf r}_2,{\bf r}_1')\rho_0({\bf r}_3)\nonumber\\
&-&H^{ex}(r_{23})\rho_0({\bf r}_2,{\bf r}_3)\rho_0({\bf r}_3,{\bf 
r}_1)\bigr]\rho_0({\bf r}_2,{\bf r}_1')\Gamma({\bf r}_1,{\bf r}_1',{\bf 
r}_2,{\bf r}_3)\Bigr\}
\label{roFSIL}
\end{eqnarray}

\begin{eqnarray}
\rho_{FSI}^{S,U}({\bf r}_1,{\bf r}_1')=\Phi ({\bf r}_1,{\bf r}_1')^A&\times&
4\rho_0({\bf r}_1,{\bf r}_1')\int d{\bf r}_2d{\bf r}_3\Bigl\{ 
\bigl[4H^{dir}(r_{23})
\rho_0({\bf r}_2)\rho_0({\bf r}_3)\nonumber\\
&-&H^{ex}(r_{23})\rho_0({\bf r}_2,{\bf r}_3)^2\bigr]\Gamma({\bf r}_1,{\bf 
r}_1',{\bf r}_2,{\bf r}_3)\Bigr\}
\label{roFSIU}
\end{eqnarray}
where $\Gamma({\bf r}_1,{\bf r}_1',{\bf r}_j)$ and $\Gamma({\bf r}_1,{\bf r}_1',{\bf r}_2,{\bf r}_3)$ denotes the product 
of the Glauber factors $G$ appearing in Eqs. (\ref{b1tilde}), (\ref{b2tildeL}) and
(\ref{b2tildeU}) minus $1$ (see Eq.(\ref{elementi}) below), and  the superscripts $S$ 
and $H$ and $L$ and $U$ stand for {\it spectator} and {\it hole}, 
and {\it linked} and {\it unlinked}, respectively.

Let us now discuss the meaning of the various terms appearing in Eq. (\ref{rodifinale}).
The first term represents the trivial shell-model contribution whereas 
$\rho_{ISC}^{H(S)}$ represents the contribution from 
initial-state correlations (ISC). If only these three contributions are 
considered the correlated momentum distribution calculated in several papers 
(\cite {bohigas}, \cite{co}) are obtained i.e. 

\begin{equation}
n({\bf k})={(2 \pi)^{-3}} \int e^{i {\bf k}({\bf r} -{\bf r}')}\rho_1 ({\bf 
r},{\bf r}') d{\bf r} d{\bf r}'
  \label{nisc}
   \end{equation}
where

\begin{equation}
\rho_1 ({\bf r}_1,{\bf r}_1') \equiv \rho_{SM}({\bf r}_1,{\bf r}_1') 
+\rho_{ISC}^H({\bf r}_1,{\bf r}_1') +\rho_{ISC}^S({\bf r}_1,{\bf r}_1') 
  \label{ro1}
   \end{equation}
As it will be clear later on using a digrammatic representation, $\rho_{ISC}^H({\bf r}_1,{\bf r}_1')$ represents the contribution when particle $"1"$  is  correlated with a second particle, whereas $\rho_{ISC}^S({\bf r}_1,{\bf r}_1')$ represents the contribution from the correlation in a spectator pair composed of particles $"2"$ and $"3"$.
The last three terms of Eq.(\ref{rodifinale}) represents the contribution from 
ISC {\it and} FSI, namely: $\rho_{ISC}^{SM}$ represents the contribution when ISC are present 
but a struck proton interacts in the final state with uncorrelated nucleons, 
whereas
$\rho_{ISC}^{H(S)}$ represents the contributions when initial state correlations 
are present but  the struck nucleon interacts either with a partner, correlated 
nucleon ($\rho_{ISC}^{H}$), or with a nucleon which is correlated with a third 
one ($\rho_{ISC}^{S}$). By taking the Forier transform of Eq. (\ref{rodifinale}) the distorted momentum distribution is obtained
\begin{equation}
n_D({\bf k}_m)={(2 \pi)^{-3}} \int e^{i {\bf k}_m({\bf r} -{\bf r}')}\rho_D 
({\bf r},{\bf r}') d{\bf r} d{\bf r}'
  \label{nd1}
   \end{equation}

Eq. \ref{ro1}, clearly illustrates the number conserving property of the expansion; as a a matter of fact, it can be readily  checked that when ${\bf r}_1={\bf r}_1'$, the integral over ${\bf r}_1$ of 
$\rho_{ISC}^H({\bf r}_1,{\bf r}_1)$ and  $\rho_{ISC}^S({\bf r}_1,{\bf r}_1)$
are identical and of opposite sign, so that the number of particles is conserved; such a property holds to all orders of the expansion.

\begin{figure}
\vspace{8 cm}
\centerline{
\epsfysize=10cm \epsfbox{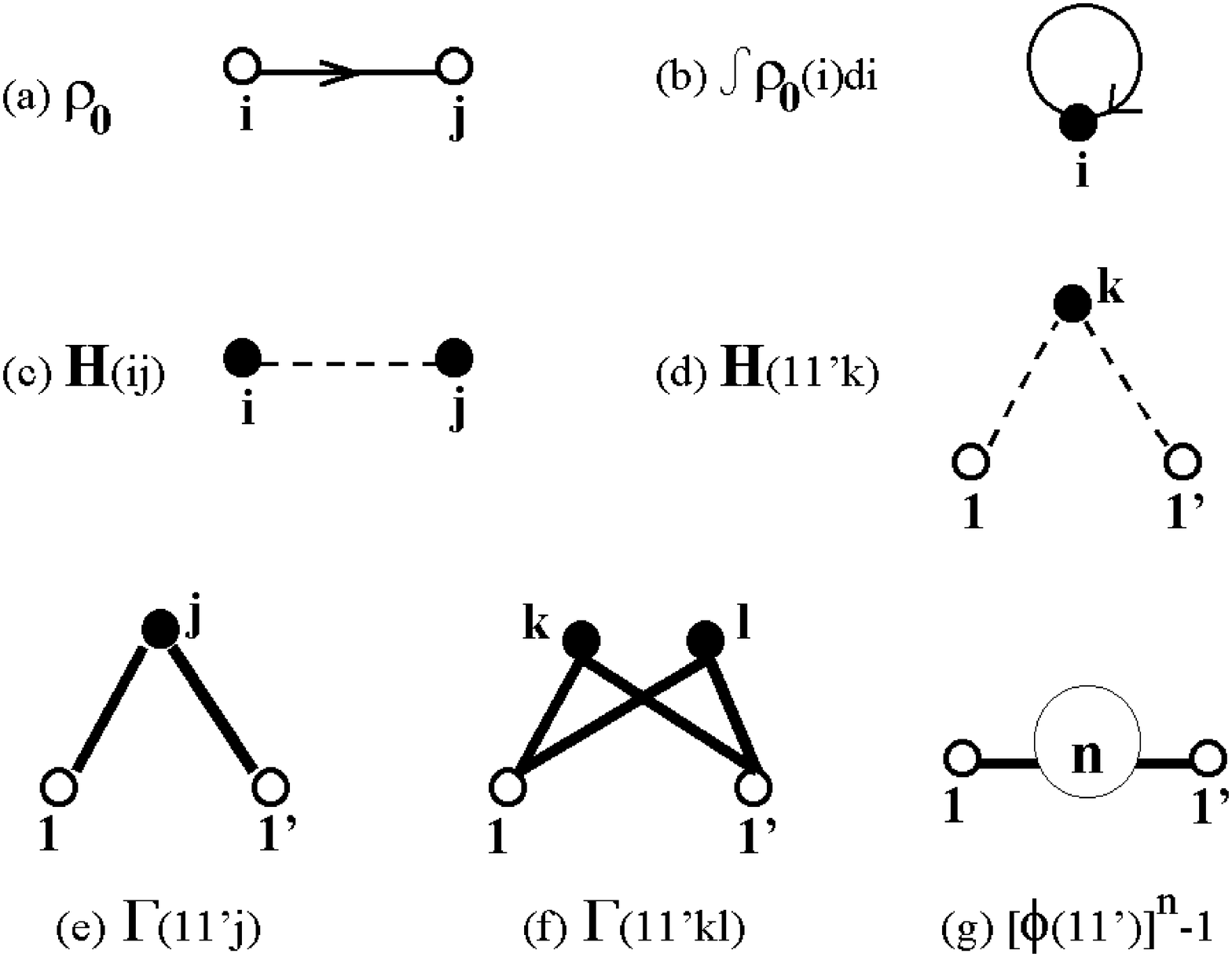}  
}
\vspace{2 cm}
\caption[ ]{ The various diagrams corresponding to the terms in Eq. (\ref{elementi}) 
 }
\label{diagrammi1}
\end{figure}

\begin{figure}
\vspace{8 cm}
\centerline{
\epsfysize=10cm \epsfbox{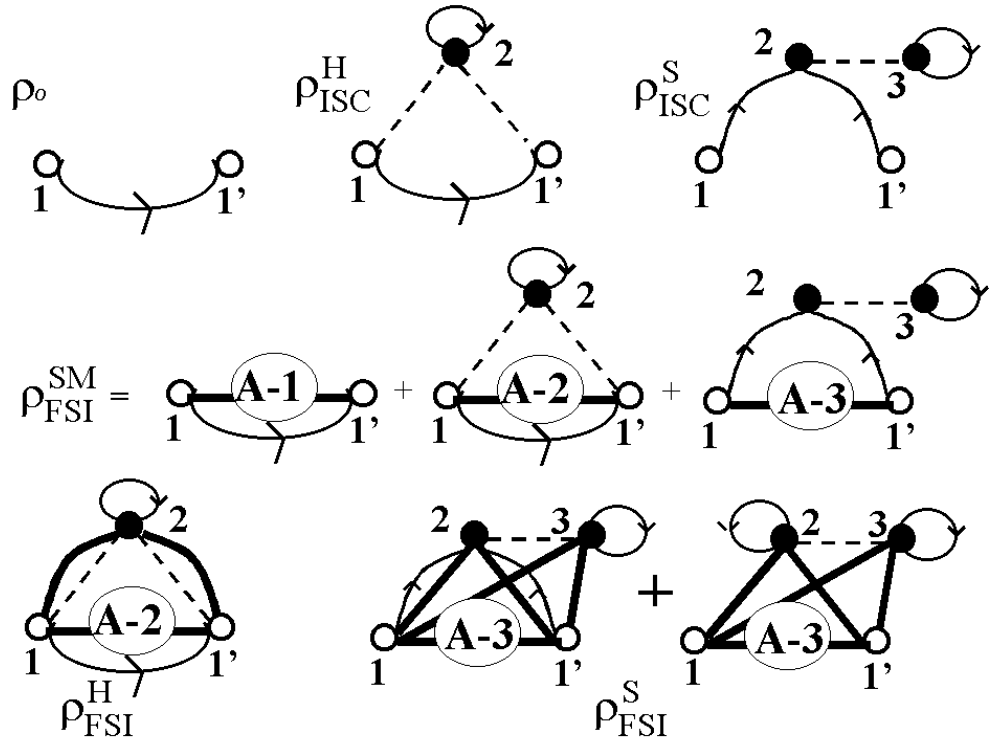}  
}
\vspace{2 cm}
\caption[ ]{ The various diagrams corresponding to the terms in Eq.
(\ref{rodifinale}) (only the direct contributions are shown) }
\label{diagrammi2}
\end{figure}

A transparent diagrammatic representation of Eq. (\ref{rodifinale}) can be given representing the generalization of the one given in \cite {ripka}-\cite {benhar} for the (undistorted) momentum distributions.
The basic elements appearing in Eq. (\ref{rodifinale}) are the following 
ones

\begin{eqnarray}
&(a)& \qquad\rho_{SM} ({\bf r}_i,{\bf r}_j)\equiv 4\rho_o({\bf r}_i,{\bf 
r}_j)\nonumber\\
&(b)& \qquad\int \rho_{SM}({\bf r}_i)d{\bf r}_i=4\int \rho_o({\bf r}_i)d{\bf 
r}_i\nonumber\\
&(c)& \qquad{\it H}^{dir(ex)}(r_{ij})\nonumber\\
&(d)& \qquad{\it H}^{dir(ex)}(r_{1k}r_{1'k})\label{elementi}\\
&(e)& \qquad\Gamma({\bf r}_1,{\bf r}_1',{\bf r}_j)\equiv G^{\dagger}({\bf 
r}_1,{\bf r}_j)G({\bf r}_1',{\bf r}_j)-1\nonumber\\ 
&(f)& \qquad\Gamma({\bf r}_1,{\bf r}_1',{\bf r}_k,{\bf r}_l)\equiv 
G^{\dagger}({\bf r}_1,{\bf r}_k)G({\bf r}_1',{\bf r}_k)G^{\dagger}({\bf 
r}_1,{\bf r}_l)G({\bf r}_1',{\bf r}_l)-1\nonumber\\
&(g)& \qquad\big [\Phi ({\bf r}_1,{\bf r}_1')\big]^n\equiv\Big [\int \rho_o({\bf 
r}_j) G^{\dagger}({\bf r}_1,{\bf r}_j)G({\bf r}_1',{\bf r}_j)d{\bf 
r}_j\Big]^n.\nonumber
\label{elementi}
  \end{eqnarray}
The diagrammatic representation of the various quantities defined in 
Eq.(\ref{elementi}) are presented in  Fig. \ref{diagrammi1}, whereas the 
diagrammatic representation of Eq.(\ref{rodifinale}) is given in Fig. 
\ref{diagrammi2}, where only the direct terms are shown. The diagrams 
corresponding to the exchange terms can be readily drawn.

\section{The nuclear transparency for $^{16}O$ and $^{40}${\bf\it 
C\lowercase{a}}}

In this section the  results of the calculation of the nuclear transparency of $^{16}O$ 
and $^{40}Ca$ obtained using Eq.(\ref{rodifinale}) will be presented. The results for the momentum distributions will be given in a separate paper
\cite {energia}.

\subsection{The nuclear wave function}

The nuclear wave function, Eq.(\ref{psi}), was constructed with $\Psi_0$ built
up from harmonic oscillator orbitals and the correlation operators
corresponding to the $V6$ Reid soft core (RSC) interaction i.e. 
$O_1(ij)=1$, 
$O_2(ij)=\mbox{\boldmath$\sigma$}_i\cdot\mbox{\boldmath$\sigma$}_j$,
$O_3(ij)=\mbox{\boldmath$\tau$}_i\cdot\mbox{\boldmath$\tau$}_j$,
$O_4(ij)=\mbox{\boldmath$\sigma$}_i\cdot\mbox{\boldmath$\sigma$}_j\mbox
{\boldmath$\tau$}_i\cdot\mbox{\boldmath$\tau$}_j$,
$O_5(ij)=S_{ij}$, 
$O_6(ij)=S_{ij}\mbox{\boldmath$\tau$}_i\cdot\mbox{\boldmath$\tau$}_j$, where 
$S_{ij}=3\bigl[(\mbox{\boldmath$\sigma$}_i\cdot{\bf 
r}_{ij})(\mbox{\boldmath$\sigma$}_j\cdot{\bf
r}_{ij})\bigr]/(r_{ij})^2-\mbox{\boldmath$\sigma$}_i\cdot\mbox{\boldmath$\sigma$}_j$.

The harmonic oscillator length parameter and the form of the
correlation functions $f_n(r_{ij})$ have been obtained by minimizing
the expectation value of the hamiltonian calculated at the second
order in the cluster expansion. The results will be presented
elsewhere\cite{energia}. Having fixed the form of the various $f_n's$
the quantities $H^{dir(ex)}$ can be readily  obtained. In the
case of the simple Jastrow wave function one has 

\begin{eqnarray}
&H&^{dir}(r_{ij})=H^{ex}(r_{ij})=f_1(r_{ij})^2-1\nonumber\\
&H&^{dir}(r_{ij},r_{i'j})=H^{ex}(r_{ij},r_{i'j})=f_1(r_{ij})f_1(r_{i'j})-1
\label{Hjastrow}
\end{eqnarray}
but when the spin, isospin and tensor dependences of the correlation functions is considered, a complex structure of $H^{dir(ex)}$ is generated.
The expressions of $H^{dir(ex)}$ for the general case of the $V6$ RSC
interaction are rather involved and will be reported elsewhere \cite {energia};
herebelow the results corresponding to the case of the dominant
correlation functions of the $V6$ RSC interaction, i.e. $f_1$, $f_4$
and $f_6$, are shown
\begin{eqnarray}
H^{dir}(r_{ij})&=&f_1(r_{ij})^2-1+27g(r_{ij})^2\nonumber\\
H^{ex}(r_{ij})&=&f_1(r_{ij})^2-1-27g(r_{ij})^2+18f_1(r_{ij})g(r_{ij})\nonumber\\
H^{dir}(r_{ij},r_{i'j})&=&f_1(r_{ij})f_1(r_{i'j})-1+27g(r_{ij})g(r_{i'j})\nonumber\\
H^{ex}(r_{ij,}r_{i'j})&=&f_1(r_{ij})f_1(r_{i'j})-1-27g(r_{ij})g(r_{i'j})+9f_1(r_
{ij})g(r_{i'j})+9f_1(r_{i'j})g(r_{ij})\nonumber\\
\label{HV6}
\end{eqnarray}
where we have used $f_4=f_6\equiv g$.

\subsection{The nuclear transparency for $^{16}O$ and $^{40}Ca$}

\begin{figure}
\vspace{8 cm}
\centerline{
\epsfysize=10cm \epsfbox{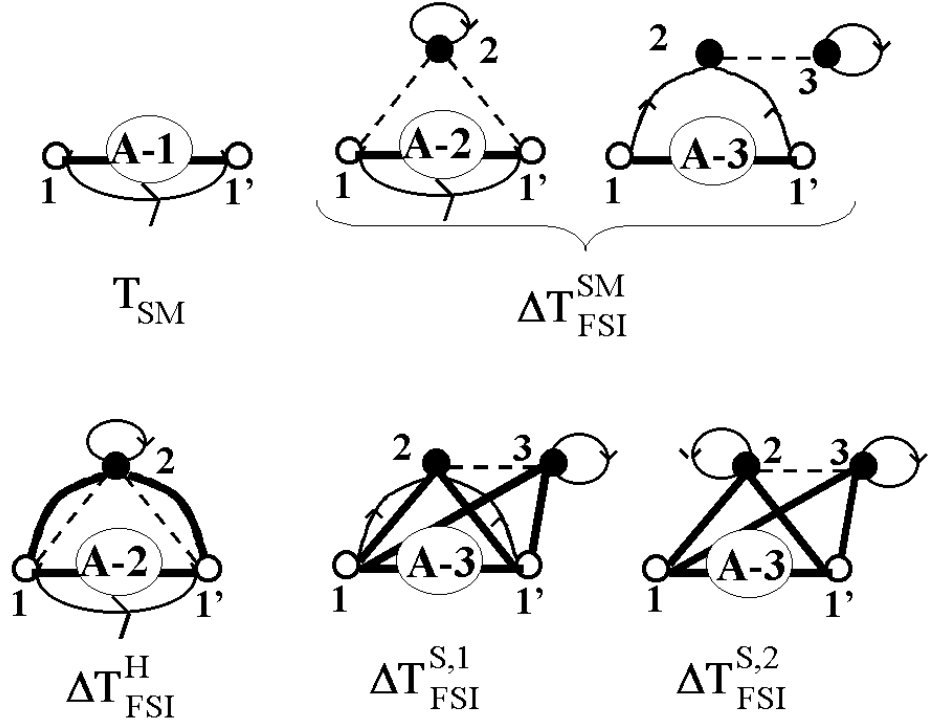}  
}
\vspace{2 cm}
\caption[ ]{ The various diagrams corresponding to the terms of the nuclear
transparency, Eq.  (\ref{trasp}) (only the direct contributions are shown) }
\label{diagrammi3}
\end{figure}

The nuclear transparency has been calculated by Eq. (\ref{ti}). Note that since 
the linked cluster expansion we are using is a number conserving one, the terms 
$\rho_{ISC}^H$ and   $\rho_{ISC}^S$ give  equal and opposite contributions to 
the integral in Eq. (\ref{ti}), so that $\Delta T$ gets contribution only from 
the terms $\rho_{FSI}^{SM}$, $\rho_{FSI}^H$, and   $\rho_{FSI}^S$; 
therefore, the nuclear transparency can be represented in the following 
form 

\begin{eqnarray}
T= 1
+ \Delta {T^{SM}_{FSI}} + \Delta {T^{H}_{FSI}} + \Delta {T^{S,1}_{FSI}} +\Delta 
{T^{S,2}_{FSI}},
  \label{trasp}
   \end{eqnarray}
where  the spectator contribution has been split in two parts which, as will 
be seen later on, cancel to a large extent. Let us reiterate that $\Delta 
{T^{SM}_{FSI}}$ includes Glauber FSI to all order between the hit nucleon and 
uncorrelated nucleons.
The diagrammatic representation of Eq. (\ref{trasp}) is given in  
Fig. \ref{diagrammi3}. Calculations have been performed by parametrising the 
Glauber profile in the usual way \cite {nikolayev} 

\begin{eqnarray}
\mit{\Gamma} (b) = \frac{\sigma_{tot}(1-i\alpha)}{4\pi b_o^2}e^{- b^2/(2b_o^2)}
  \label{gamma}
   \end{eqnarray}
with $\sigma_{tot} =43\ mb$, $\alpha = -0.33$ and $b_o = 0.6\ fm$. Two different types of   
nuclear wave functions have been used, viz.  the wave function, Eq.(\ref{psi}), corresponding to 
the Reid V6 interaction \cite{vuotto},  with single particle and correlation 
parameters determined from the minimisation of the nuclear hamiltonian 
\cite{benhar}, and the phenomenological Jastrow wave function with central correlations,
frequently used in the calculations of the transparency (see e.g. 
\cite{bianconi}).  The results of the calculations, which  are presented in  
Table \ref{table1} and \ref{table2}, 
deserve 
the following comments:
\begin{enumerate}
\item Within the phenomenological central correlation approach, the effects of 
correlations on the nuclear transparency  is sizeable (about 12\%)
\item The contribution of the spectator term  is almost zero, originating from 
two terms of opposite sign, and the effect of FSI  within correlated nucleons is 
almost entirely due to the hole contribution
\item Non-central correlations affect very sharply the nuclear transparency, in 
that the overall effect of correlations reduces to about 2\%, with the hole 
contribution remaining the dominant one and the spectator contribution canceling 
out.
\end{enumerate}
 
It is important to stress that similar conclusions have been reached in \cite {cmt1},
where the nuclear transparency in the process $^4He(e,e'p)X$ 
has been obtained by an exact (to all order of correlations and Glauber multiple scattering) calculation 
performed using a realistic four-body wave function corresponding to the 
same interaction used in this paper.

Thus we have found a small effect  of realistic correlations  on the transparency, in apparent agreement
 with the results of, e.g., Ref. \cite{nikolayev}; there, however, such a result is a consequence of a cancellation between hole and spectator contributions, whereas in our approach it is due to an overall decrease of the transparency generated by non central correlations, which lead  to an almost vanishing contribution of the spectator effect, with  the only surviving contributions being $\Delta T_{FSI}^{SM}$ and $\Delta T_{FSI}^{H}$ \footnote{Note that in the  central Jastrow correlation approach, both for complex nuclei (cf. Table 1) and for $^4He$ ( cf. \cite{bianco} and \cite{energia}, where the Jastrow calculation has been carried out to all orders  both in the correlations and the multiple scattering operators), correlations increase the transparency by more than 10\%.}. The reasons of the apparent overall  agreement between our results and the ones of Ref.\cite{nikolayev}, are, at the moment,  difficult to understand, also in view of the fact that the two approaches are formally diffferent, with the latter one being based upon the Foldy-Walecka expansion \cite{foldy}, which requires the orthonormality condition $\int d{\bf r}_1\rho ({\bf r}_1)C({\bf r}_1,{\bf r}_2) = 0$, which, however, is not usually  implemented  in actual calculations.

\section{Summary and Conclusions}

Our work can be summarised as follows:
\begin{enumerate}
\item A linked cluster expansion has been developed which includes both 
initial state correlations and final state interactions. The expansion holds for the most general form of the correlation function, which includes both central and non-central correlations, and 
 is such that at each order in the correlations, Glauber multiple scattering is included at all orders. 
\item The expansion has been applied to the calculation of the nuclear transparency 
in the processes $^{16}O(e,e'p)X$ and $^{40}Ca(e,e'p)X$. The results show that 
 whereas  central Jastrow correlations increase the transparency
 by about 12\%,  realistic central and non
central correlations increase it by only 2\%.
\item A comparison of our results with the ones obtained for the nuclear transparency
in the process  $^{4}He(e,e'p)X$ calculated by an exact treatment of realistic correlations and Glauber multiple scattering (\cite {energia} , \cite{cmt1}) show similar results, indicating that the effects of correlations on triple- and higher order Glauber multiple scattering contributions is neglegible. A thorough investigation of the convergence of the distorted linked cluster expansion will be presented elsewhere \cite {energia}, together with the results of the calculations for the distorted momentum distributions.
\end{enumerate}

To sum up, the general conclusion can be drawn that a realistic calculation of the nuclear transparency in semi-inclusive processes $A(e,e'p)X$, for both light and heavy nuclei, can be performed, thus appreciably improving the pioneering estimates based on simple phenomenological nucler wave functions embodying only central repusive correlations.

\section {Acknowledgments} 
We are indebted to Hiko Morita and Kolya Nikolaev for many useful discussions.

\begin{table}
\caption[dummu6]{The nuclear transparency, Eq. (\ref{trasp}), for $^{16}O$  .} 
\begin{flushleft}
\renewcommand{\arraystretch}{1.2}
\begin{tabular}{llllllllllll}  
\hline\noalign{\smallskip}
   &   \qquad  $T_{SM}   \qquad \qquad  $  &  $\Delta {T^{SM}_{FSI}}$ 
\qquad\qquad &   $\Delta {T^{H}_{FSI}}$ \qquad\qquad &  $\Delta{T^{S,1}_{FSI}}$ 
\qquad\qquad & 
$\Delta{T^{S,2}_{FSI}}$ \qquad\qquad& $T$  \qquad\qquad \\
\hline
Central    & \qquad 0.51 & 0.020 & 0.032  & --0.013 & 0.022 & 0.57 \\
Realistic  & \qquad 0.51 & 0.003  & 0.009 & 0.001 & --0.001 & 0.52 \\
\noalign{\smallskip}\hline
\end{tabular}
\renewcommand{\arraystretch}{1}
\label{table1}
\end{flushleft}
\end{table} 
\begin{table}
\caption[dummu6]{The nuclear transparency, Eq. (\ref{trasp}), for $^{40}Ca$ .} 
\begin{flushleft}
\renewcommand{\arraystretch}{1.2}
\begin{tabular}{llllllllllll}  
\hline\noalign{\smallskip}
   &   \qquad  $T_{SM}   \qquad \qquad  $  &  $\Delta {T^{SM}_{FSI}}$ 
\qquad\qquad &   $\Delta {T^{H}_{FSI}}$ \qquad\qquad &  $\Delta{T^{S,1}_{FSI}}$ 
\qquad\qquad & 
$\Delta{T^{S,2}_{FSI}}$ \qquad\qquad& $T$  \qquad\qquad \\
\hline
Central    & \qquad 0.41 & 0.020 & 0.028  & --0.011 & 0.023 & 0.47 \\
Realistic  & \qquad 0.41 & 0.002  & 0.008 & --0.001 & 0.001 & 0.42 \\
\noalign{\smallskip}\hline
\end{tabular}
\renewcommand{\arraystretch}{1}
\label{table2}
\end{flushleft}
\end{table} 



\end{document}